\documentstyle[11pt]{article}

\begin{document} 
\centerline{\bf Quantization of a one-dimensional time-dependent periodic system with  }
\centerline{\bf Hamiltonian and constants of motion approaches}
\vskip2pc
\centerline{G. L\'opez}
\centerline{Departamento de F\'{\i}sica, Universidad de Guadalajara}
\centerline{Apartado Postal  4-137, 44410 Guadalajara, Jalisco, M\'exico}
\vskip2pc
\centerline{PACS: 03.65.-w, 03.65.Ca, 03.65Sq}
\vskip2cm
\centerline{ABSTRACT}
\vskip1pc\noindent
For a particle moving in a one-dimensional space an under a periodic external force, its quantization is
study using the Hamiltonian (generalized linear momentum quantization) and constant of motion
(velocity quantization) approaches. it is shown a great difference on the quantization of both approaches and
the ambiguities arisen  by using the quantization on the constants of motion.
\vfil\eject\noindent  
{\bf 1. Introduction}
\vskip0.5pc\noindent
Notwithstanding the great success of the Hamiltonian and Lagrangian [1] approaches to quantize a classical 
dynamical system [2], there are still some dynamical systems which their consistent quantization can be
questioned [3,4].  In particular, the so called dissipative systems [5] and time-explicitly depending
systems [6] present some problems for their consistent formulation. One of the main problems that a
one-dimensional dissipative system presents is that most of the times it is not possible to have explicitly
the velocity  in terms of the generalized linear momentum and position, $v=v(x,p)$, from the usual
definition of the generalized linear momentum $p=\partial L/\partial v$,  where $L$ is the associated
Lagrangian of the system. In turns, this means that the Hamiltonian associated to this system,
$H=vp-L(x,v)$, can not be given explicitly and remains implicit within the associated constant of motion
[7], $K=K(x,v)$. It is shown [8] that a consistent quantization for this type of systems can be gotten by
using the association of an Hermitian operator to the velocity variable,
$\hat{v}=-i(\hbar/m)\partial/\partial x$, the constant of motion, $\hat{K}(x,\hat{v})$, and proceeding to
solve the associated Shr\"odinger equation,
$$i\hbar{\partial\Psi\over\partial t}=\widehat{K}(x,\hat{v})\Psi\ ,\eqno(1)$$
where $\Psi=\Psi(x,t)$ is the wave function. The question arises whether or not this same approach can be
extended to nonautonomous systems (explicitly time dependent systems). In this paper, a one-dimensional
time-dependent  periodic dynamical system is studied to see whether or not a natural consistent extension
of Eq. (1) can be achieved for this system. In the first part, the Hamiltonian quantization approach is
used and its known natural ambiguity is pointed out. In the second part, three classical constants of
motion are deduced with units of energy and their quantization is carried out  through the extension of Eq.
(1). Finally, the ambiguity resulting from this quantization approach  is pointed out.
\vskip2pc
\leftline{\bf 2. Hamiltonian quantization approach}
\vskip1pc\noindent
Consider the following nonautonomous dynamical system
$$\dot x=p/m\ ,\hskip3pc \dot p=-A\cos{\omega t}\ ,\eqno(2)$$
where $p$ is the generalized linear momentum, $x$ represents the position of the particle, $m$ is the mass
of the particle, $\omega$ and $A$ are the angular frequency and amplitude of the external oscillating
force. A Hamiltonian associated to this system is given by
$$H={p^2\over 2m}=xA\cos{\omega t}\ .\eqno(3a)$$
Note that one gets the following limits
$$\lim_{A\to 0}H={p^2\over 2m}\hskip3pc\hbox{and}\hskip3pc\lim_{\omega\to 0}H={p^2\over 2m}+Ax\eqno(3b)$$
which are what one could expect with the corresponding limits in Eq. (2). Furthermore, one could add an
arbitrary time dependent function to Eq. (3a)  and still having  a Hamiltonian which the Hamilton
equations satisfy Eq. (2). In fact, this is true for any nonautonomous dynamical system. Then, one can say
that two Hamiltonian
$H$ and
$H'$ are equivalent if they are related by the following expression
$$H'=H+f(t)\ ,\eqno(4)$$
where $f(t)$ is an arbitrary function. This defines an equivalence relation [10] where a nonautonomous
dynamical system is characterized by a class of Hamiltonians
$$[H]=\biggl\{H' (Hamiltonian)|~H'=H+f(t)\biggr\}\ .\eqno(5)$$
This ambiguity, however, does not affect the classical dynamics  neither the quantum dynamics, as one will
sees below. 
\vskip0.5pc\noindent
To quantize our system, one needs to solve the associated Shr\"odinger equation
$$i\hbar{\partial\Psi\over\partial t}=\widehat{H}(x,\hat p)\Psi\ ,\eqno(6)$$
where $\Psi=\Psi(x,t)$ is the wave function, $\widehat{H}$ and $\hat p=i\hbar\partial/\partial x$ are the
Hermitian operators associated to the classical variables $H$ and $p$. Consider also the expansion of the
wave function $\Psi$ in terms of the basis $\{|k\rangle\}_{k\in\Re}$,
$$\Psi(x,t)=\int dk~C(k,t)|k\rangle\ ,\eqno(7)$$
where $k$ is given by $k=\sqrt{2mE_k/\hbar^2}$, being $E_k$ the energy of the free particle. The basis
$\{|k\rangle\}_{k\in\Re}$ has the following relations
$$\langle x|k\rangle={e^{ikx}\over\sqrt{2\pi}}\ ,\hskip2pc \langle k'|k\rangle=\delta(k-k')\ ,\
\hskip2pc \hat{p}|k\rangle=\hbar k|k\rangle\eqno(8a)$$
with the braket $\langle k'|k\rangle$ defined as
$$\langle k'|k\rangle=\int dx~\langle k'|x\rangle\langle x|k\rangle\ .\eqno(8b)$$
Using (8b) and (3a), substituting (7) in (6), and multiplying to the left by the bra $\langle k'|$, one gets
the a first order partial differential equation for $C$
$$i\hbar C(k',t)=E_{k'}C(k',t)+iA\cos{\omega t}{\partial C(k',t)\over\partial k'}\ .\eqno(9)$$
One can simplify this equation doing the following change of variable
$$C(k,t)=e^{-iE_kt/\hbar}D(k,t)\ .\eqno(10)$$
The resulting equation for the $D's$ coefficients is
$${\partial D\over \partial t}-{A\over\hbar}\cos{\omega t}{\partial D\over \partial k}=0\eqno(11)$$
which can be solved by the characteristics method [9]. Given the initial condition $C(k,0)=D(k,0)=F(k)$, one
has
$$D(k,t)=F\biggl(k+{A\over\hbar\omega}\sin{\omega t}\biggr)\ .\eqno(12)$$
Of course, the initial condition $C(k,0)=F(k)$ is related with the initial wave function
$\Psi_0(x)=\Psi(x,0)$ as $\Psi_o(x)=\int dk~F(k)|k\rangle$. That is, $F(k)$ is just the Fourier
transformation of $\Psi_0(x)$.
 Thus, the coefficients $C's$ are given by
$$C(k,t)=F\biggl(k+{A\over\hbar\omega}\sin{\omega t}\biggr)~e^{-i\hbar k^2t/2m}\ .\eqno(13)$$
With this expression, the solution of Eq. (6) is fully determined. Now, the ambiguity (4) will be
reflected in Eq. (13) by an additional phase, $-(i/\hbar)\int f(t)dt$ which does not depend on the variable
$k$, and this arbitrary phase corresponds to an arbitrary local element of the unitary group $U(1)$.
However, since $|\Psi(x,t)|^2$ gives the probability to find the particle in the position $x$ at the time
$t$, this global phase has not contribution on the quantum dynamics of the system.
\vskip3pc\noindent
\leftline{\bf 3. Constants of motion quantization approach}
\vskip0.5pc\noindent
The nonautonomous dynamical system (2) can be written in terms of the position, $x$, and velocity $v$,
variables as
$$\dot x=v\ ,\hskip3pc\dot v=-{A\over m}\cos{\omega t}\ ,\eqno(14)$$
and a constant of motion for this system is a function $K=K(x,v,t)$ such that $dK/dt=0$. That is, it
satisfies the following first order partial differential equation
$$v{\partial K\over\partial x}-{A\over m}\cos{\omega t}{\partial K\over\partial v}+{\partial K\over\partial
t}=0\ .\eqno(15)$$
This equation can be solved by the characteristics method [9], bringing about the following characteristics
curves
$$C_1=v+{A\over m\omega}\sin{\omega t}\eqno(16a)$$
and
$$C_2=t\biggl(v+{A\over m\omega}\sin{\omega t}\biggr)+{A\over m\omega^2}(\cos{\omega t}-1)-x\ .\eqno(16b)$$
Therefore, the general solution of Eq. (15) is given by
$$K=G(C_1,C_2)\ ,\eqno(17)$$
where $G$ is an arbitrary function of both characteristics curves. Let us note that $C_1$ and $C_2$ have
the following limits 
$$\lim_{\omega\to 0}C_1=v+{At\over m}\ ,\hskip2pc\lim_{\omega\to 0}C_2=t\biggl(v+{At\over m}\biggr)
-{At^2\over 2m}-x\eqno(18a)$$
and
$$\lim_{A\to 0}C_1=v\ ,\hskip2pc\lim_{A\to 0}C_2=vt-x\ .\eqno(18b)$$
Now, there are at least three different ways to get constants of motion with units of energy. One could
select the function $G$ as $G=(m/2)C_1^2$ to get the constant of motion
$$K_1(x,v,t)={1\over 2}mv^2+{vA\over\omega}\sin{\omega t}+{A\over 2m\omega^2}\sin^2{\omega t}\ .\eqno(19)$$
Other way of selecting $G$ is of the form $G=(m/2)C_1^2-AC_2$. This selection brings about the following
constant of motion
\begin{eqnarray*}
K_2(x,v,t)&=&{m\over 2}\biggl(v^2+{2vA\over m\omega}\sin{\omega t}+{A^2\over m^2\omega^2}\sin^2{\omega
t}\biggr)\\ & &-Atv-{A^2t\over m\omega}\sin{\omega t}-{A^2\over m\omega^2}(cos{\omega t}-1)+Ax\ .
\end{eqnarray*}
$$\eqno(20)$$
In addition, one could select $G$ of the form $G=(m\omega/2)C_1C_2$ to get the constant of motion
\begin{eqnarray*}
K_3(x,v,t)&=&{m\omega t\over 2}\biggl(v^2+{2vA\over m\omega}\sin{\omega t}+
{A^2\over m^2\omega^2}\sin^2{\omega t}\biggr)+
{Av\over m\omega}(\cos{\omega t}-1)\\ & &+{A^2\over 2m\omega^2}\sin{\omega t}(cos{\omega t}-1)
-{m\omega\over 2}\biggl(xv+{Ax\over m\omega}\sin{\omega t}\biggr)\ .
\end{eqnarray*}
$$\eqno(21)$$
These constants of motion have units of energy and have the following limits
$$\lim_{\omega\to 0}K_1={m\over 2}\biggl(v-{At\over m}\biggr)^2\ ,\hskip2pc\lim_{A\to 0}K_1={1\over
2}mv^2\ ,\eqno(22a)$$
$$\lim_{\omega\to 0}K_2={1\over 2}mv^2+Ax ,\hskip2pc\lim_{A\to 0}K_2={1\over 2}mv^2\ ,\eqno(22b)$$
and
$$\lim_{\omega\to 0}K_3=0 ,\hskip2pc\lim_{A\to 0}K_3=0\ .\eqno(22c)$$
The quantization of the system (14) through the constants of motion will be carried out with the
association of an Hermitian operator to the velocity, $\hat{v}=-(i\hbar/m)\partial/\partial x$, and the
constant of motion, $\widehat{K}(x,\hat{v})$. Then, one will proceed to solve the Shr\"odinger equation
$$i\hbar{\partial\Psi\over\partial t}=\widehat{K}(x,\hat{v},t)\Psi\ .\eqno(23a)$$
Note that within this velocity quantization approach, the Heisenberg's uncertainty relation is expressed as
$$\Delta x\Delta v\ge{\hbar\over m}\ ,\eqno(23b)$$
and the following relations are gotten straightforwardly
$$[x,x]=[\hat{v},\hat{v}]=0\,\hskip2pc [x,\hat{v}]={i\hbar\over m}\ ,\hskip1pc\hbox{and}\hskip1pc
\hat{v}|k\rangle={\hbar k\over m}|k\rangle\ ,\eqno(24)$$
where $[,]$ represents the commutator of two operators, and the state $|k\rangle$ is given by (8a). Now,
proposing the wave function of the form (7), multiplying to the left by the bra $|k'\rangle$, and using
the above properties, one gets for the above three constants of motion the following equations associated to
their  coefficients (renaming $k'$ by $k$)
$$i\hbar\dot{C}^{(1)}(k,t)=\biggl[{\hbar^2 k^2\over 2m}+{A\hbar k\over m\omega}\sin{\omega t}+
{A\over 2m\omega^2}\sin^2{\omega t}\biggr]C^{(1)}(k,t)\ ,\eqno(25a)$$
$$i\hbar\dot{C}^{(2)}(k,t)=B(k,t)C^{(2)}(k,t)+iA{\partial C^{(2)}(k,t)\over\partial k}\ ,\eqno(25b)$$
and
$$i\hbar\dot{C}^{(3)}(k,t)=f(k,t)C^{(3)}(k,t)+{i\hbar\omega\over 4}C^{(3)}(k,t)-{im\omega\over 2}
\biggl[{\hbar k\over m}+{A\over m\omega}\sin{\omega t}\biggr]{\partial C^{(3)}(k,t)\over\partial k}\
,\eqno(25c)$$
where the functions $B(k,t)$ and $f(k,t)$ have been defined as
\begin{eqnarray*}
B(k,t)&=&{\hbar^2 k^2\over 2m}+{A\hbar k\over m\omega}\sin{\omega t}+{A\over 2m \omega^2}\sin^2{\omega t}\\
& &-{A\hbar kt\over m}-{A^2t\over m\omega}\sin{\omega t}+{A^2\over\omega^2}(1-\cos{\omega t}) 
\end{eqnarray*}
$$\eqno(26a)$$
and
\begin{eqnarray*}
f(k,t)&=&{m\omega t\over 2}\biggl[{\hbar^2 k^2\over m^2}+{2A\hbar k\over m^2\omega}\sin{\omega t}+
{A^2\over m^2\omega^2}\sin^2{\omega t}\biggr]\\
& &+{A\hbar k\over 2m\omega}(\cos{\omega t}-1)+{A^2\over 2m\omega^2}\sin{\omega t}(\cos{\omega t}-1)\ .
\end{eqnarray*}
$$\eqno(26b)$$
Note that in order to associate an Hermitian operator to the constant of motion $K_3$, one had to make use
of the Weyl's quantization to the product $xv$, $\widehat{xv}=(x\hat{v}+\hat{v}x)/2$. Eqs. (25's) represent
first order partial differential equations, and given the initial condition $C^{(i)}(k,0)=F(k)$ for
$i=1,2,3$, their solutions are given by
$$C^{(1)}(k,t)=F(k)e^{-i\phi_1(k,t)}\ ,\eqno(27a)$$
$$C^{(2)}(k,t)=F\biggl(k+{At\over\hbar}\biggr)e^{-i\phi_2(k,t)}\ ,\eqno(27b)$$
and
$$C^{(3)}(k,t)=F\biggl(ke^{-\omega t/2}+g(t)-g(0)\biggr)e^{\omega t/4-i\phi_3(k,t)}\ ,\eqno(27c)$$
where the phases $\phi_1$, $\phi_2$, and $\phi_3$ and the function $g(t)$ have been defined as
$$\phi_1(k,t)={\hbar k^2t\over 2m}-{kA\over m\omega^2}(\cos{\omega t}-1)+{A\over 2m\omega^2}
\biggl({t^2\over 2}-{\sin{2\omega t}\over 4\omega}\biggr)\ ,\eqno(28a)$$
$$\phi_2(k,t)={1\over\hbar}\int_0^tB\biggl(k+{At\over\hbar}-{As\over\hbar},s\biggr)~ds\ ,\eqno(28b)$$
$$\phi_3(k,t)={1\over\hbar}\int_0^tf\biggl(e^{\omega s/2}(k-e^{-\omega t/2}+g(t)-g(s)),s\biggr)~ds\
,\eqno(28c)$$
and
$$g(t)={A\over 5\hbar\omega}(2\cos{\omega t}+A\sin{\omega t})\ .\eqno(28d)$$
Note from Eq. (27c) that the real exponent has to be canceled with the negative exponent coming from the
initial conditions to keep the probability finite at any time. On the other hand, the probabilities
$|C^{(i)}(k,t)|^2$ for i=1,2,3 tell us quite different behavior of the quantum system. The solution (27a)
tell us essentially that the quantum system described by the constant of motion (19) will remain in the
same initial quantum state but with a complicated phase. Solution (17b) tell us that the quantum system
will change linearly its state of momentum $k$ with respect the time. Finally, the solution (27c) tell us
that the quantum system described by the constant of motion (21) will oscillate with respect the time in a
form determined by (28d). 
\vskip0.5pc\noindent
These ambiguities about selecting a proper constant of motion to quantize the dynamical nonautonomous
system (14) make the approach of using (23a) a little bite subtle. 
\vskip3pc\noindent
\leftline{\bf 4. Conclusions and Comments}
\vskip0.5pc\noindent
The study of Hamilton and constants of motion quantization approaches has been made for a one-dimensional
nonautonomous periodic dynamical system. It was shown that both approaches brings about different solutions
for the quantized system. In addition, it is not clear how to select a proper constant of motion to have
the same solution as the Hamiltonian approach has, Eq. (13). At first sight, one might think that since the
limits (3b) and (22b) are consistent, their associated quantum solutions (13) and (27b) should be also 
consistent (look alike). However, this is not the case. Therefore, a possible extension of Eq. (1)
(autonomous systems) to Eq. (23a) (nonautonomous systems) is a little bite subtle.
\vskip5pc\noindent
\leftline{\bf Acknowledgements}
\vskip0.5pc\noindent
 This work was supported by SEP under the contract PROMEP/103.5/04/1911 and the University of Guadalajara.

\vfil\eject
\leftline{\bf References}
\obeylines{
1. H. Goldstein, {\it Classical Mechanics}, Addison-Wesley, Reading,MA,1950.
\quad R.M. Santilli, {\it Foundations of Theoretical Mechanics, Vol. I ,II},
\quad Springer-Verlag, Berlin/New York, 1978.
2. A. Messiah, {\it Quantum Mechanics,, Vol. I,II}, 
\quad North-Holland, Amsterdam 1961.
\quad P.A.M. Dirac, {\it The Principles of Quantum Mechanics}, 
\quad Clarendon Press Oxford, 1992.
\quad R.P. Feynman and A.R. Hibbs, {\it Quantum Mechanics and Path Integrals},
\quad McGraw-Hill, London, New York, 1965. 
3. G. L\'opez, Rev. Mex. F\'{\i}s.,{\bf 48} (2002) 10.
4. G. L\'opez and G. Gonz\'alez, Int. Jour. Theo. Phys.,{\bf 43},10 (2004) 1999.
5. P. Havas, Suppl. Nuovo Cimento {\bf 5}, 10 (1957) 363.
\quad H. Dekker, {\it Classical and Quantum Mechanics of the Damped Harmonic
\quad Oscillator}, North-Holland, Amsterdam, 1981.
\quad R. Glauber and V.I. Man'ko, Sov. Phys. JEPT {\bf 60}, (1984) 450.
\quad V.V. Dodonov et al, Hadron J. {\bf 4} (1981) 173.
\quad G. L\'opez, Ann. Phys. {\bf 251} (1996) 372.
\quad V.E. Tarasov, Phys. Lett. A, {\bf 288} (2001) 173.
6. G. L\'opez, IL Nuovo cimento B, {\bf 115} (2000) 137.
\quad G. L\'opez and J.I Hern\'andez, Ann. Phys., {\bf 193} (1989) 1.
7. G. L\'opez, L.A. Barrera, Y. Garibo, H. Hern\'andez, J.C. Salazar and 
\quad C.A. Vargas, Int. Jour. Theo. Phys., {\bf 43},10 (2004) 2009. 
8. See reference 4.
9. F. John, {\it Partial Differential Equations}, Springer-Verlag, 
\quad Berlin/New York, 1974.
\quad G. L\'opez, {\it Partial Differential Equations of First Order and Their 
\quad Applications to Physics}, World Scientific, 1999.
10. I.N. Herstein, {\it Topics in Algebra}, Xerox College Publishing, 
\quad 1964. Chapter 1.
}

\end{document}